\documentclass{ragtime} 
\usepackage{times}

\title[Backflow in simulated MHD accretion disks]%
      {Backflow in simulated MHD accretion disks}

\author[R. Mishra, 
        M. \v{C}emelji\'{c}    
        and W\l odek Klu\'{z}niak]
       {Ruchi Mishra\at{1,a} 
        Miljenko \v{C}emelji\'{c}\at[]{1,2} 
        and W\l odek Klu\'{z}niak\at[]{1}\\
        \ins{1}Nicolaus Copernicus Astronomical Center, Polish academy of Sciences,\\
        Bartycka 18, 00-716
Warsaw, Poland\splitins[1]
       \\
        \ins{2}Academia Sinica, Institute of Astronomy and Astrophysics, P.O. Box 23-141, \splitins[2]Taipei 106, Taiwan\\
        \\
        \ins{a}\Email{rmishra@camk.edu.pl}}

\coentry{Z. Stuchl\'{\i}k, J. Kov\'a\v{r} and J. Cimrman}%
       {Equilibrium of spinning test particles in equatorial plane\par
        of Kerr--de~Sitter spacetimes}

\begin{document}

\begin{abstract}
 We perform resistive MHD simulations of accretion disk with alpha-viscosity, accreting onto a rotating star endowed with a magnetic dipole. We find backflow in the presence of strong magnetic field and large resistivity, and probe for the dependence on Prandtl number. We find that in the magnetic case the distance from the star at which backflow begins, the stagnation radius, is different than in the hydrodynamic case, and the backflow shows non-stationary behavior. We compare the results with hydrodynamics simulations.
\end{abstract}


\section{Introduction}\label{intro}
 When matter falls onto a  massive object, it often takes the form of a rotating gaseous disk, known as an accretion disk. The process of accretion is understood by a mechanism where the angular momentum from gas is transported outwards which allows the matter to slowly fall into the central object. Accretion plays important role in formation of most astronomical objects such as galaxies, stars and planets. Hence it is important to understand the accretion process.
 
 We are particularly interested in a special class of objects consisting of magnetized stellar type objects, such as T-Tauri stars and white dwarfs or neutron stars in close binary systems. They are mostly surrounded by accretion disk, and have a well defined magnetosphere. In order to understand how accretion takes place in presence of magnetic field we perform non-ideal magneto-hydrodynamic (MHD) numerical simulations.

 While spanning the parameter space, we find that in some cases the accretion flow is directed away from the central star. We explore conditions for such behaviour. After the Introduction, In \S 2 we briefly present our numerical setup, and in \S 3 present results of our simulations. In \S 4 we present a comparison with the purely hydrodynamical (HD) case.

\section{Numerical setup}
We perform two-dimensional axisymmetric, viscous and resistive magnetohydrodynamic star-disk simulations. Details of our setup are presented in \cite{cem19}. We use the publicly available PLUTO code (v.4.1) \cite{m07,m12}, with a logarithmically stretched grid in radial direction in spherical coordinates, and uniformly spaced co-latitudinal grid. Resolution is $R\times\theta=[217\times100]$ grid cells, stretching the domain to 30 stellar radii, in a quadrant of the meridional plane. The solved equations are, in CGS units:
 \begin{equation}
     \frac{\partial \rho}{\partial t} +\nabla.(\rho \vec{v})= 0
 \end{equation}
 \begin{equation}
     \frac{\partial 
(\rho \vec{ v})}{\partial t} +\nabla.\bigg[\rho \vec{v}\vec{v}+\big(P{+\frac{B^2}{8\pi}}\big)\widetilde{I}-{\frac{\vec{B}\vec{B}}{8\pi}}-\widetilde{\tau}\bigg]= \rho g
 \end{equation}
 
 \begin{equation}
    \frac{\partial E}{\partial t} +\nabla\cdot\bigg[(E+P+\frac{B^2}{8\pi})\vec{v}-\frac{(\vec{v}\cdot\vec{B})\vec{B}}{4\pi}\bigg]=\rho\vec{g}\cdot\vec{v}
 \end{equation}

 \begin{equation}
     {\frac{\partial \vec{B}}{\partial t} +\nabla\times(\vec{B}\times\vec{ v}+\eta_m\vec{J})= 0}
 \end{equation}
The above equations are continuity equation, momentum equation, energy equation and induction equation respectively. The symbols have their usual meaning: $\rho$ and $v$ are the matter density and velocity, $P$ is the pressure, $B$ is the magnetic field and $\eta_{\rm m}$ and $\widetilde{\tau}$ represent the resistivity and the viscous stress tensor, respectively.

We perform a parameter study by changing the magnetic field strength, resistivity and alpha viscosity for star rotating at 10$\%$ of the equatorial mass-shedding limit $\Omega=0.1\ \Omega_{\mathrm {br}}$. In our setup the viscosity parameter $\alpha_{\mathrm v}$, which describes the strength of the  viscous torque, allowing the disk to accrete, is varied from 0 to 1. The resistivity parameter $\alpha_{\mathrm m}$, which defines coupling of the stellar magnetic field with the disk material, is also varied from 0 to 1. The effect of changes in those two parameters is described by the magnetic Prandtl number:
\begin{equation}
    P_\mathrm{m}= \frac{2}{3}\frac{\alpha_\mathrm{v}}{\alpha_\mathrm{m}}
\end{equation}
Each simulation in our parameter study is run until a quasi-stationary state is reached.
 \begin{figure*}
   \centering
   \includegraphics[width=\hsize]{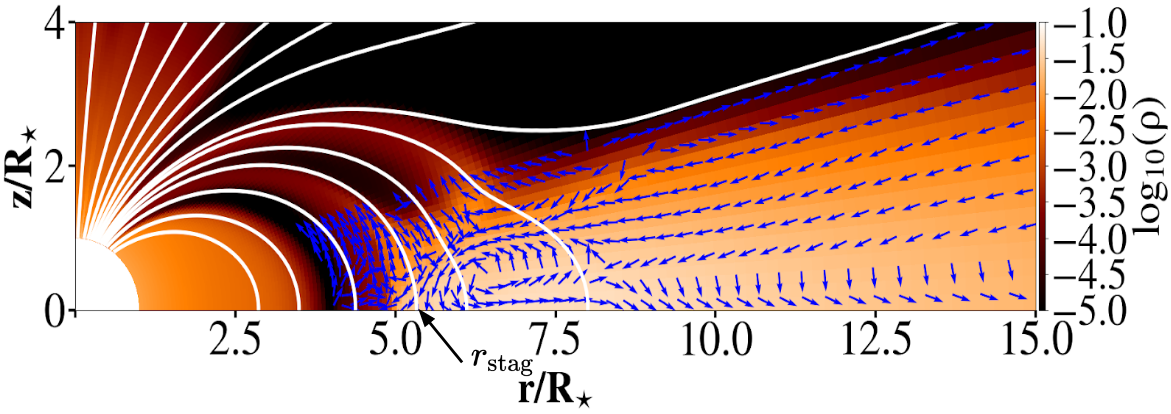}
      \caption{Density in a logarithmic colour grading in MHD simulation with $\alpha_{\mathrm v}=1$, $\alpha_{\mathrm m}=0.4$, with the initially dipolar magnetic field of $1000~G$, for a slowly rotating star. The stagnation radius $r_{\mathrm{stag}}=5.5$ stellar radii. The blue arrows indicate the direction of flow.}  
         \label{FigVibStab}
   \end{figure*} 

\section{Backflow in MHD disk}
A snapshot in a quasi-stationary state in our simulation is shown in Fig.~\ref{FigVibStab}. The accretion disk is truncated at a few stellar radii due to magnetic pressure. The accretion flow is channelled into a funnel flow where the flow follows the magnetic field lines. In the outer region of the disk the accretion flow is towards the central star. In the inner region of the disk, along the midplane of the disk, the flow is away from the star--this flow is termed as backflow in the accretion disk. Such equatorial outflow was first reported in the analytical work by  \cite{urpinb}. Global solution for three dimensional viscous alpha accretion disk was obtained in \cite{Kita95} and \cite{KK00}. Equatorial backflow is not uncommon. It was obtained in numerical simulations in \cite{kl92}, \cite{igum95}, \cite{roz}, and also recently in MRI simulations by \cite{bm} and in GRMHD simulation of RIAF (Radioactively inefficient accretion flow) in \cite{white}\\ 
\begin{table*}
\centering
 \begin{tabular}{||c c c c c c c c||} 
 \hline
 $\alpha_{\mathrm v}$ & $\alpha_{\mathrm m}$ & $P_{\mathrm m}=\frac{2}{3}\frac{\alpha_{\mathrm v}}{\alpha_{\mathrm m}}$ & $B_\star$(G) & $\Omega_{\star}/\tilde{\Omega}$& Backflow  & $R_{\mathrm{stag}}$ &Type\\ [0.5ex] 
 \hline\hline

0.1 & 0.1 & 0.60 & 1000 & 0.1 & No & -- & --\\

 \hline

 0.4 & 0.1 & 2.60 & 1000 & 0.1 & No& -- & -- \\
 
 \hline

 1.0 & 0.1 & 6.60 & 1000 & 0.1 & No & -- & --\\
 
 \hline

 0.1 & 0.4 & 0.16 & 1000 & 0.1 & No & -- & --\\
 
 \hline

0.4 & 0.4 & 0.60 & 1000 & 0.1 & No & -- & -- \\
\hline
 
1.0 & 0.4 & 1.60 & 1000 & 0.1 & No & -- & -- \\

\hline

 0.1 & 1.0 & 0.06 & 1000 & 0.1 & Yes & $6\pm1$ & steady \\
 \hline
 
  0.2 & 1.0 & 0.13 & 1000 & 0.1 & Yes & $7.5\pm1.5$ & steady \\
  \hline
  0.3 & 1.0 & 0.20 & 1000 & 0.1 & Yes & $6\pm1$ & steady \\
  \hline
 
 0.4 & 1.0 & 0.26 & 1000 & 0.1 & Yes & $6\pm 2$ & steady\\
 \hline
 
  0.5 & 1.0&  0.30  & 1000 & 0.1 & Yes & $6\pm1$ & steady \\
  \hline
  
  0.6 & 1 0 &  0.40  & 1000 & 0.1 & Yes & $6\pm1$ & steady \\
  
  \hline
  
  0.7 & 1.0 &  0.46  & 1000 & 0.1 & Yes & $8\pm0.5$ &  intermittent\\
  \hline
   
  0.8 & 1.0 & 0.53 & 1000 & 0.1 & Yes & $8\pm0.5$ &  intermittent\\
  \hline
 
   1.0 & 1.0 & 0.60 & 1000 & 0.1 & No & -- & -- \\
   \hline
 \end{tabular}
 \caption{\label{Table 1}List of parameters and results : presence of backflow, stagnation radius and type of backflow for strongly magnetized, slowly rotating stars}
\end{table*}
Backflow appears in the disk for particular combinations of $\alpha_{\mathrm v}$ and $\alpha_\mathrm m$ parameters. In the Table~1 we present different parameters, and we check whether there is backflow in the disk. Lower values of resistive parameter $\alpha_\mathrm m$ restrict backflow in the disk. For higher values of resistive parameter, a backflow is obtained. We obtain two kinds of backflows: a steady, or an intermittent flow. Intermittent backflow is found as we approach higher values of viscosity parameter. We also show the presence of backflow in dependence on magnetic Prandtl number $P_{\mathrm m}$. Above a critical value of magnetic Prandtl number, which is about $P_{\mathrm m}\sim 0.6$, there is no backflow in the disk.

\section{Comparison with backflow in HD disk}

 \begin{figure*}
   \centering
   \includegraphics[width=\hsize]{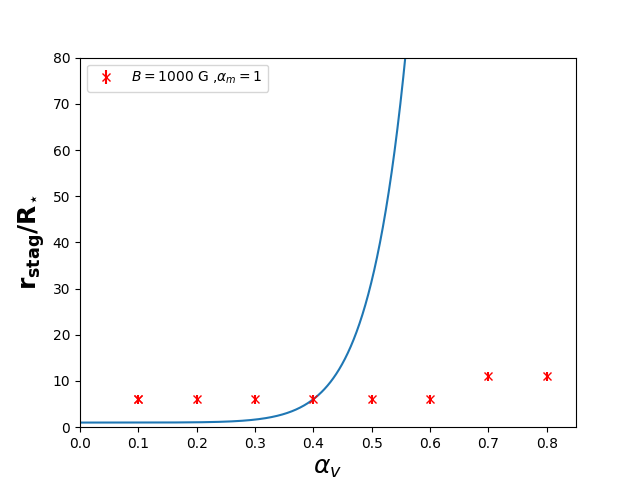}
      \caption{Position of the stagnation radius in simulations with different viscosity coefficients $\alpha_\mathrm v$ for magnetic resistive coefficient $\alpha_\mathrm m=1$. The blue solid curve is from the \cite{KK00} purely HD analytical solution.}
         \label{mag}
   \end{figure*}

Initializing with the \cite{KK00} solutions, we performed numerical simulations in purely hydrodynamical cases in \cite{RM}. Backflow is obtained in the mid-plane of the disk for alpha viscosity coefficient $\alpha_\mathrm v < 0.6$. The starting point of backflow, stagnation radius, is found to be a function of alpha viscosity. In our MHD simulations we obtain backflow for even higher values of $\alpha_\mathrm v$. The stagnation radius, in the case of slowly rotating stars, shows to be independent of alpha viscosity--see Fig.~\ref{mag}.

\section{Conclusions}
We find backflow in the simulated MHD disk in a part of the parameter space. In the presence of strong magnetic field and high resistivity, we obtain backflow for higher values of viscous parameter than critical $\alpha_\mathrm v$ in purely HD case. We find a dependence of backflow on magnetic Prandtl number, $P_\mathrm m$, where for values $P_\mathrm m<0.6$ there is a backflow. As we approach the critical $P_\mathrm m$, there is intermittent backflow in the disk. We do not find the same relationship of stagnation radius and viscosity parameter as the one obtained in purely HD cases.

\ack

 Work at the Copernicus Center was funded by the Polish NCN grant No. 2019/33/B/ST9/01564. M\v{C} developed the setup for star-disc simulations while in CEA, Saclay, under the ANR Toupies grant, and partly worked on it while in Shanghai Astronomical Observatory, supported by CAS President’s International Fellowship for Visiting Scientists (grant No. 2020VMC0002). We thank ASIAA/TIARA (PL and XL clusters) in Taipei, Taiwan and NCAC (CHUCK cluster) in Warsaw, Poland, for access to Linux computer clusters used for high-performance computations. We thank the {\sc pluto} team for the possibility to use the code.

\bibliography{ragsamp}
\end{document}